# Life Cycle Sustainability Assessment of Fossil- and Bio-Based Membrane Manufacturing

*Elizabeth Aigaje-Espinosa, Christina Carbrello[†], Rui Shi[‡,]**

Department of Chemical Engineering, The Pennsylvania State University, University Park, PA, 16801.

**Corresponding author: rms6987@psu.edu*

KEYWORDS

ultrafiltration, polymeric membranes, hollow fibers, life cycle assessment, environmental impact, biopolymers, uncertainty analysis.

ABSTRACT

The transition towards sustainable membrane manufacturing is challenged by its reliance on fossil-based polymers, organic solvents, and generation of large volumes of contaminated water. We combined process simulation, life cycle assessment, and Monte Carlo simulation to identify hotspots in the manufacturing of ultrafiltration hollow-fiber membranes for pharmaceutical applications and evaluate strategies to mitigate impact and reduce waste generation. Process models informed by industry input compared fossil- and bio-based membranes on a cradle-to-gate basis per square meter of membrane produced. Key environmental hotspots include polymer and bore solution preparation, driven by the contribution of polymer and solvent, accounting for 30-80% of total impacts across the categories investigated, and wastewater dilution, driven by the environmental burden of solvent-contaminated wastewater management. Under the modeled conditions, although polymer substitution can reduce environmental burdens, cellulose-based membranes show higher impacts than polysulfone membranes in at least 85% of simulations within each impact category, due to chemical-intensive cellulose modification and regeneration. In contrast, solvent recycling can mitigate solvent- and wastewater-related burdens when combined with reduced water demand, yielding climate change benefits at water demands below approximately 36 L per $m^2$ under the modeled recovery configuration and highlighting its potential to reduce the environmental footprint of membrane manufacturing.

## 1. Introduction

Membrane-based separation technology is widely used across industrial sectors, including water and wastewater treatment, pharmaceuticals, food processing, chemical, and petrochemical industries[1]. Its advantages over other separation processes include compact design, mild operating conditions, relatively simple operation, and the ability to perform separations under conditions that are difficult for other techniques[2]. In addition, membranes are often regarded as a clean and environmentally friendly option due to low energy and chemical requirements and minimal waste generation during operation[3,4].

However, despite these operational advantages, membrane manufacturing remains resource-intensive, relying heavily on fossil-based polymers and toxic solvents, while generating significant volumes of contaminated wastewater[3,5]. As industries increasingly seek to reduce dependence on fossil resources and environmental impacts, there is a growing need to address the sustainability of membrane production itself, ensuring that upstream manufacturing aligns with the environmental benefits associated with membrane operation. In addition, several solvents and non-solvents used in membrane fabrication have been identified as posing risks to human health[6]. In practice, these chemicals are often transferred to wastewater streams, which in many cases are discharged without adequate treatment[5]. Therefore, improving manufacturing sustainability is not only an environmental necessity but also relevant for reducing potential human health risks.

Sustainability assessment tools such as life cycle assessment (LCA) offer a framework for quantifying environmental and human health impacts across the life cycle of products and processes[7]. By identifying environmental hotspots, LCA enables the development of targeted strategies that reduce the environmental footprint of transforming activities. To date, LCA studies of membrane manufacturing have primarily explored two approaches to reduce environmental impacts: (i) replacing fossil-based polymers with bio-based alternatives and (ii) substituting toxic solvents with less hazardous or lower-impact options[8–11]. Among bio-based polymers, cellulose acetate (CA) has received the greatest attention and remains the only alternative that has demonstrated practical substitutability at industrial scales due to its favorable mechanical properties, chemical stability, and relatively low cost[12,13]. Previous studies report that replacing polysulfone (PSF) with CA can reduce contributions to human health, ecosystem quality, and resource depletion[8]. However, these benefits are not consistent across all impact categories; reductions in global warming and fossil resource scarcity remain limited due to the chemical-intensive acetylation process required for CA production[9]. Beyond CA, other bio-based candidates, such as polylactic acid and chitosan, have been proposed as membrane materials for applications such as water treatment and gas separation[14–18]. Nevertheless, their implementation remains largely limited to laboratory-scale studies due to challenges associated with membrane performance, stability, and scalability. In addition to polymer substitution, green solvents have emerged as an alternative for reducing the environmental, health, and safety concerns associated with conventional solvents used in membrane manufacturing, particularly N-methyl-2-pyrrolidone (NMP). Several lower-impact alternative solvents have been proposed in the membrane literature, including bio-based options such as methyl lactate (ML), ethyl lactate (EL), Cyrene™, and less-toxic and biodegradable alternatives including organic carbonates, Rhodiasolv® PolarClean, and dimethyl sulfoxide (DMSO)[19–22]. While these alternatives often exhibit lower environmental impacts or reduced toxicity at the solvent level, their implementation can modify membrane fabrication conditions, material requirements, and production yields. In some cases, these process changes have been shown to outweigh the environmental benefits associated with green solvents, resulting in higher membrane-level impacts[4,23].

The above LCA studies have provided valuable insights into the environmental impacts of membrane manufacturing; nonetheless, most rely on simplified process representations. While informative, these approaches do not capture detailed process models and material and energy flows, limiting the ability to identify environmental hotspots and evaluate opportunities to mitigate impacts. In addition, current assessments often overlook the process adjustments required when replacing materials, despite evidence that such changes can substantially influence overall environmental performance. Another important gap concerns wastewater management. Membrane manufacturing generates large volumes of wastewater streams that require treatment and/or dilution before discharge due to the presence of solvents such as NMP[5,24,25]. Although wastewater generation is frequently identified as an environmental concern in membrane manufacturing, the influence of wastewater generation, dilution, or treatment requirements on overall sustainability has received limited attention. As a result, the interactions between material selection, process operation, wastewater management, and environmental performance remain insufficiently understood.

To address these gaps, this study integrates process simulation, LCA, and uncertainty and scenario analyses to capture process-level details, with model assumptions informed and validated through collaboration with industry partners. Specifically, we develop process models for fossil- and bio-based ultrafiltration hollow-fiber membranes for pharmaceutical applications, compare their environmental performance, identify environmental hotspots, and evaluate strategies to reduce waste generation and environmental impacts. This approach provides practice-oriented insights and supports material selection and process design decisions to advance sustainability in the membrane industry.

## 2. Methods

### 2.1 Membrane manufacturing process

In this study, the manufacturing process model targets ultrafiltration (UF) hollow fiber membranes produced via non-solvent induced phase separation (NIPS), one of the most widely used techniques for polymeric membrane fabrication[19,20,26]. UF hollow fiber membranes were selected based on their broad implementation in industrial separations, including pharmaceutical processing[26]. The manufacturing process encompasses six main processing areas (Figure 1 and Figure S1): (1) polymer & bore solution preparation, (2) extrusion & coagulation, (3) rinsing, (4) conditioning & drying, (5) module assembly, and (6) wastewater dilution.

This process configuration represents a general NIPS-based manufacturing framework applicable to polymeric hollow-fiber membranes, including PSF and related polymers such as polyethersulfone (PES), using polar aprotic solvents (e.g., NMP, DMSO, dimethylacetamide (DMAc)) and pore-forming additives (e.g., polyethylene glycol (PEG), polyvinylpyrrolidone (PVP)). For the baseline system evaluated in this study, PSF, NMP, and PEG were selected as the polymer, solvent, and pore-forming additive, respectively. The corresponding baseline process conditions for each stage, summarized in Table 1 and Table S6, were established based on the operating ranges (Table S15) compiled from industry input and literature sources. These baseline conditions represent the specific configuration modeled in this study and may vary with the polymer, solvent, and additive selection, as well as the targeted membrane properties.

In *polymer & bore solution preparation*, the polymer solution is prepared by dissolving polysulfone (PSF) in N-methyl-2-pyrrolidone (NMP), with polyethylene glycol (PEG) added as a pore-forming agent[27]. The bore solution consists of a water/NMP mixture; the presence of solvent in the bore moderates the rate of phase inversion, enabling more controlled lumen formation[28]. Both solutions are degassed before extrusion. Then, in *extrusion & coagulation*, the polymer and bore solutions are co-extruded through the spinneret to form nascent hollow fibers. The fibers are immersed in a coagulation bath, where phase inversion is induced using deionized (DI) water as the non-solvent, following a *rinsing* stage to remove residual NMP from the fibers. In *conditioning & drying*, the nascent fibers are immersed in a glycerol conditioning bath to minimize pore collapse and shrinkage during subsequent drying[28]. Finally, in the *module assembly* stage, the dried hollow fibers are bundled and introduced into membrane cartridges to produce finished membrane modules.

The coagulation, rinsing, and conditioning baths operate as continuous-flow systems, with fresh water or conditioning solutions fed continuously to prevent NMP accumulation and maintain the target glycerol concentration. Due to the limited availability of studies quantifying water consumption during coagulation and rinsing, a water demand of 150 kg $m^{-2}$ membrane was assumed for each bath, resulting in a total baseline water demand of 300 kg $m^{-2}$ membrane. This estimate was derived from reported average wastewater generation values for membrane fabrication processes[5,25], and assigned to the coagulation and rinsing stages, as these are the most water-intensive operations in the process. The resulting effluent streams, which contain NMP as the primary contaminant, are assumed to be discharged to an external wastewater treatment (WWT) facility, consistent with common industrial practice. Before discharge, an on-site *wastewater dilution* step is included to ensure compatibility with conventional treatment systems. Because specific discharge limits for NMP are not widely established in the United States[29], chemical oxygen demand (COD) was used as a surrogate parameter to estimate the dilution factor for NMP-containing effluents. COD is a widely used indicator of the total oxidizable organic load in wastewater and therefore provides an indirect measure of the contribution of NMP and other organic constituents to the wastewater burden (e.g., ethanol). Although acceptable influent COD concentrations vary among treatment plants, this study adopted the COD concentration specified in the ecoinvent[30] activity "treatment of wastewater, average, wastewater treatment" as the target concentration for the dilution stage (Table 1), ensuring consistency with the background life cycle inventory. Additional modeling details and estimation of water for dilution are provided in Section S1.

For the cellulose acetate (CA)-based membrane scenario, the manufacturing process is identical to that described above, with the addition of a *regeneration* step (Figure 1 and Figure S1). This treatment, applied to membranes intended for pharmaceutical separation applications, involves treating the CA membrane with a NaOH-ethanol (EtOH) solution, which hydrolyzes the acetate groups and converts the membrane to regenerated cellulose (RC). Ethanol used during regeneration is assumed to be recovered and recycled within the process, with makeup ethanol to compensate for losses. Following regeneration, the fibers are rinsed with DI water and subsequently undergo the same conditioning and downstream steps described for the PSF membrane (Figure 1).

**Table 1.** Material compositions and operating conditions used to simulate PSF and RC membrane manufacturing. Values represent baseline conditions selected from ranges compiled from industry input and literature.

| Process area | Parameter (unit) | Parameter value |
|---|---|---|
| Polymer & bore solution preparation | Polymer fraction | 0.18 |
| | PEG fraction | 0.10 |
| | Solvent fraction, bs | 0.5 |
| | Volume ratio: bs to ps | 1 |
| Extrusion & coagulation | Non-solvent demand (kg $m^{-2}$ membrane) | 150 |
| | Coagulation temperature (°C) | 20 |
| Regeneration[a] | NaOH concentration (M) | 0.2 |
| Rinsing | Water demand (kg $m^{-2}$ membrane) | 150 |
| | Rinsing temperature (°C) | 50 |
| Conditioning & drying | Glycerol concentration (wt. %) | 25 |
| | Glycerol pore filling fraction[b] | 1.1 |
| | Drying temperature (°C) | 50 |
| Wastewater dilution | Target COD (mg $L^{-1}$) | 425.34 |

[a]Only applicable to RC membranes. [b]Parameter used to estimate glycerol consumption based on membrane porosity. Abbreviations: bs = bore solution, ps = polymer solution. Complete parameter list is provided in Table S6.

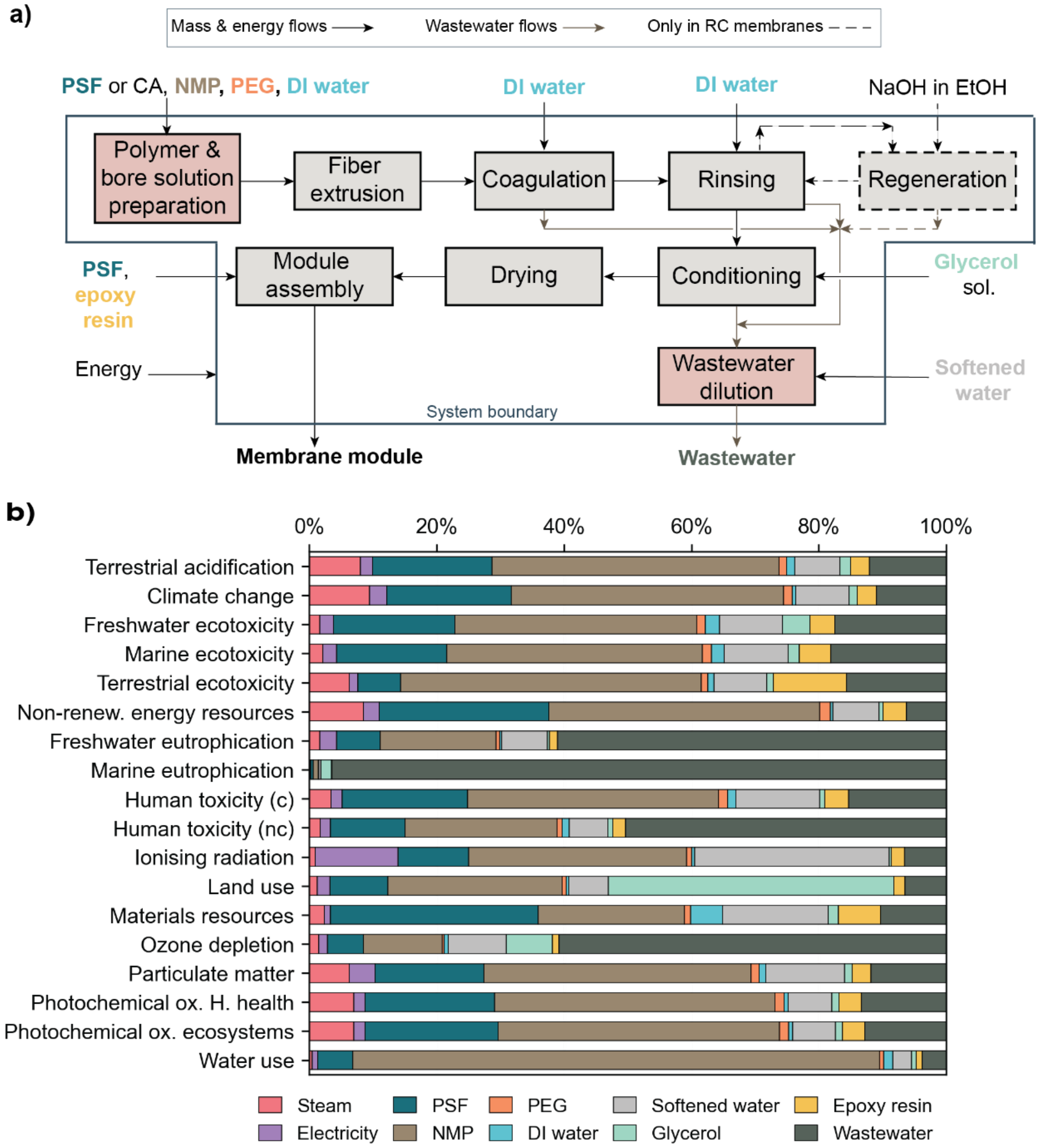

**Figure 1.** a) LCA system boundary and process block diagram for membrane manufacturing. Light, pink-colored processing areas are the environmental hotspots. b) Baseline LCA for PSF membrane production, showing the contribution breakdown across impact categories. Absolute impact values for all categories are provided in Table S11. Abbreviations: Non-renew = non-renewable, c = carcinogenic, nc = non-carcinogenic, ox = oxidation.

## 2.2 Unified framework: membrane manufacturing simulation, LCA, uncertainty, sensitivity and scenario analyses.

To translate the manufacturing process described above into process-level material and energy flows, the Cytiva UFP-300-E-85 hollow fiber module[31,32] was used as a reference for PSF and RC membrane and module specifications, with the main characteristics summarized in Table S4. These specifications were used solely to establish the relationship between polymer content, membrane area, and module throughput, providing a basis for material quantification. The baseline membrane formulation and manufacturing conditions were defined separately, as presented in Table 1. Combined with a production basis of 8000 L of polymer solution processed per day (four 2000 L polymer solution tanks), these relationships enabled the conversion of literature-reported energy and material demands, commonly provided per $m^2$ of membrane, into process-level flows for simulation. The process simulation, life cycle assessment (LCA), uncertainty quantification, global sensitivity, and scenario analyses were implemented within a unified open-source Python framework[33], leveraging QSDSan[34], BioSTEAM[35] and Brightway25[36] packages.

The LCA was conducted following ISO 14040/14044 standards[37,38], the goal was to compare the environmental impacts associated with the production of fossil-based PSF membranes and bio-based RC membranes, as well as to identify hotspots and strategies to reduce waste generation and environmental impact. The functional unit was defined as 1 $m^2$ of UF hollow fiber membrane, providing a common basis for comparing manufacturing-related environmental burdens. Regenerated cellulose membranes have been reported to exhibit permeation and antifouling characteristics comparable to fossil-derived polymeric membranes[27,39], supporting their consideration as a potential alternative for similar separation applications. Despite this reported comparability, the present study focuses on membrane production rather than equivalent separation service; therefore, differences in permeability, rejection, and operating flux were not evaluated.

The system boundary (Figure 1a) encompasses membrane manufacturing, wastewater dilution, and upstream emissions associated with energy and material inputs, resulting in a cradle-to-gate assessment. The foreground life cycle inventory (LCI) was built using the mass and energy balances obtained from simulation and from industry data; the background inventory was modeled using the ecoinvent 3.10 dataset[30]. Unless otherwise specified, chemical inputs were modeled using fossil-derived background datasets. Glycerol and ethanol in the membrane conditioning and regeneration stages, respectively, were modeled using bio-based production pathways, representing the increasing use of renewable feedstock for chemical production and their relevance to industrial decarbonization strategies. A complete list of foreground and background inventory datasets is provided in Table S12-S14. The environmental impacts were quantified using the ReCiPe 2016 v1.03, midpoint (H) methodology[40]. We adopted a cutoff approach, under which wastewater generated during the process was attributed as an environmental burden to the membrane manufacturing system.

The uncertainty analysis was performed using Latin Hypercube Sampling[41] (n=3000 samples), incorporating uncertainty from 44 technological variables (i.e., operating conditions and process parameters) and background inventory data. For the technological parameters, variability ranges and probability distributions were assigned based on data availability. Triangular distributions were used for parameters supported by industrial input or literature data, whereas uniform distributions were assigned when ranges were provided, or when supporting evidence was limited[42,43]. For the background inventory, uncertainty associated with data quality was evaluated via the pedigree matrix approach[44,45]. Additional details regarding sample size selection, as well as the complete list of variables, uncertainty ranges, and assigned probability distributions are provided in Section S2.2 of the Supplementary Information. The uncertainty model was propagated to the global sensitivity analysis (GSA), performed using Spearman's rank correlation coefficients[41] to identify the parameters with the greatest influence on each impact category. The GSA outcome was then used to guide the scenario analyses, which evaluated the role of solvent recycling, water demand, and rinsing temperature on system sustainability. For the solvent recycling scenarios, uncertainty analysis (3000 simulations) was conducted for each water demand case, comparing scenarios with and without recycling, with water demand treated as a fixed parameter. For the rinsing temperature-water demand scenarios, a baseline LCA was performed for each combination of temperature and water demand values.

For the polymer candidates (section 3.2.1), the cradle-to-gate LCA and uncertainty analysis followed the methodology described above. The system boundaries for PSF and cellulose acetate (CA) production are shown in Figure 2. The foreground LCI for CA production was collected from patents[46,47] and literature sources[8,10,48], considering wood-derived cellulose as feedstock. For PSF, the foreground inventory was built using the ecoinvent 3.10 dataset supplemented with additional energy and solvent requirements associated with solvent recovery and makeup streams[49]. Background inventories for both polymer candidates were obtained from ecoinvent 3.10. Moreover, to explore opportunities for reducing the environmental impacts of CA production, we additionally evaluated a scenario in which conventional acetic acid was replaced with bio-based acetic acid (bio-AA) produced via $CO_2$ fermentation. In this case, the background inventory for acetic acid was substituted with a literature-based inventory for bio-AA[50]. The complete life cycle inventory for all polymer production scenarios is provided in section S2.1.

**3. Results and Discussion:**

### 3.1 Baseline LCA and hotspot analysis: Fossil-based membrane

Figure 1b illustrates the contribution of energy and material flows to the environmental impacts of fossil-based PSF membrane manufacturing across the assessed impact categories. NMP is the dominant contributor in 13 of the 18 categories, accounting for 23-83% across impacts. The contribution of PSF originates from both the membrane and the module housing (see Figure 1), contributing 6-33% across most categories, except in marine eutrophication, where its contribution is negligible. Softened water used for *wastewater dilution* contributes between 3% (water use score) and 31% (ionizing radiation score), while the wastewater generated during this stage is the primary contributor to marine eutrophication, freshwater eutrophication, human toxicity (non-carcinogenic), and ozone depletion. Glycerol, employed during conditioning, dominates only the land use category (45%), primarily due to feedstock cultivation burdens associated with vegetable oil-derived glycerol production, while contributing less than 4% in all the other categories. Other materials, including PEG, DI water, and epoxy resin, contribute less than 6% individually.

Beyond material inputs, utilities represent an additional source of environmental burden. Extrusion is the most electricity-intensive operation, accounting for approximately 50% of the total electricity demand, whereas rinsing dominates thermal energy consumption, contributing nearly 83% of the total heating demand. Together, energy requirements contribute less than 15% to the overall environmental impacts across all categories, with the largest contributions observed for climate change (10%, primarily from steam) and ionizing radiation (13%, primarily from electricity).

Membrane manufacturing is a materials-intensive process, with material inputs dominating contributions across all impact categories. Two processing areas, *polymer & bore solution preparation* and *wastewater dilution*, were identified as the primary environmental hotspots, together contributing 45-98% of the total environmental burden. Within the *polymer & bore solution preparation* stage, fossil-based NMP is the dominant contributor due to its extensive use in both polymer and bore solutions. Its environmental burden also propagates downstream to the *wastewater dilution* stage. In particular, the impacts associated with softened water and wastewater streams (Figure 1) highlight the significant environmental burden associated with NMP-containing effluents. Together, these results suggest that improving the sustainability of membrane manufacturing requires an integrated approach that extends beyond replacing fossil-derived solvents and polymers. Process-oriented strategies, including solvent management and minimization of wastewater generation and treatment requirements, are necessary for achieving meaningful reductions in environmental impacts.

### 3.2. LCA under uncertainty: Bio- and fossil-based membranes

Based on the baseline analysis presented in Section 3.1, reducing the environmental burden associated with fossil-derived polymers emerges as an opportunity for improving the sustainability of membrane manufacturing. Among potential alternatives, cellulose acetate (CA) was identified as a promising candidate due to its bio-based origin, technical maturity, and suitability for bioprocessing and pharmaceutical membrane separation applications[27,51].

#### 3.2.1. LCA of (bio)polymer candidates for membrane manufacturing

To evaluate the potential environmental benefits of replacing PSF with CA, the environmental impacts associated with the production of 1 kg of each polymer were compared. PSF production follows a conventional polymerization route based on the condensation reaction between 4,4′-dichlorodiphenyl sulfone (4,4′-DCDPS) and bisphenol A in a solvent medium (NMP), followed by precipitation in a non-solvent (methanol) and subsequent purification steps (Figure 2a). The system boundary for CA production (Figure 2b) includes cellulose pulp pretreatment, acetylation, and downstream precipitation, assuming an acetic acid recovery rate of approximately 80%[8,48].

PSF exhibits a lower climate change impact, with a 5th-95th percentile range of 10.36-15.32 kg $CO_2$-eq $kg^{-1}$ polymer, compared to 13.96-16.63 kg $CO_2$-eq $kg^{-1}$ for CA. For PSF, the climate change score is primarily driven by the monomer demand in polymerization, whereas CA impact is dominated by the acetylation step (Figure 2c). Although the cellulose feedstock itself contributes negligibly to the overall score, the high demand and environmental burden associated with fossil-based acetic anhydride (~1.7 kg per kg CA)[26] and acetic acid (~0.7 kg per kg CA)[48,52] offset the potential environmental benefits of the biobased feedstock. The greater environmental burden of CA relative to PSF is consistent across all impact categories, except for material resources, where CA exhibits lower impact (Figure 2d).

To mitigate the environmental burden associated with acetylation in CA production, we evaluated replacing fossil-based acetic acid with bio-based acetic acid (bio-AA), produced via $CO_2$ fermentation. This alternative has been reported to achieve up to 70% lower carbon footprint relative to fossil-based acetic acid[50]. Incorporating bio-AA in CA production (CA (bio-AA)) reduced the climate change score of CA to 12.96-15.76 kg $CO_2$-eq $kg^{-1}$ (5th-95th percentile), yet it remained higher than that of PSF. Nevertheless, CA (bio-AA) achieved the lowest scores across several other impact categories, including ecotoxicity, non-renewable energy use, human toxicity (carcinogenic), and material resource use (Figure 2d).

Although CA (bio-AA) does not outperform PSF in terms of climate change impacts, the reductions observed in non-renewable energy demand suggest that replacing PSF with CA (bio-AA) could reduce dependence on fossil-derived energy sources in polymer production. Additionally, improvements across toxicity- and resource-related impact categories indicate broader environmental benefits and support the potential role of cellulose-based polymers in the transition toward more sustainable membrane technologies. Importantly, bio-based acetic acid is already produced at industrial scale by companies such as LanzaTech[53] in the U.S and Lenzing[54] in Europe, supporting the technical feasibility of this scenario.

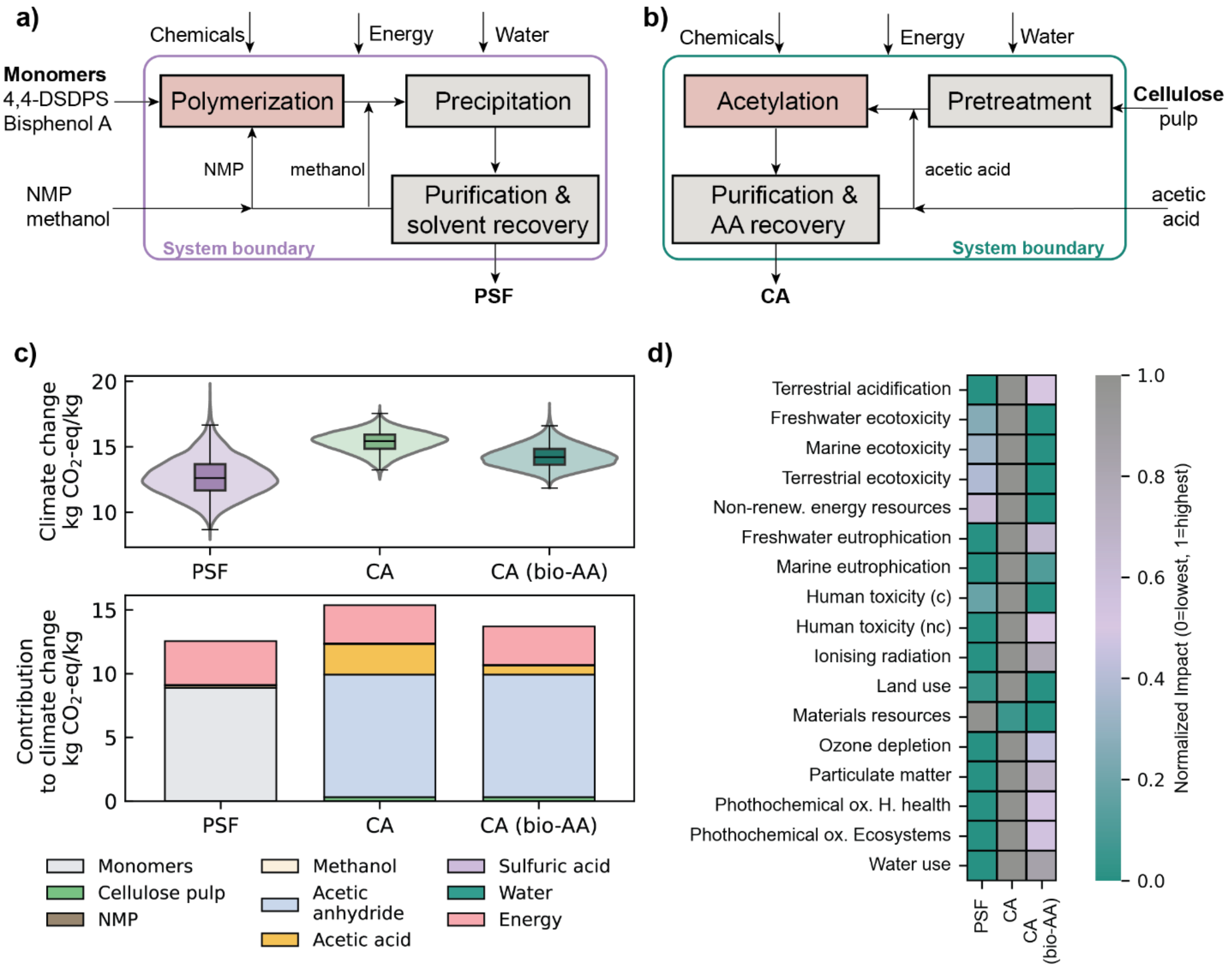

**Figure 2.** LCA of polymer candidates for membrane manufacturing. System boundaries and main production stages for a) polysulfone (PSF) and b) cellulose acetate (CA). Light, pink-colored processing areas are the environmental hotspots. c) Violin plots overlaid with box plots comparing the climate change scores under uncertainty for PSF, CA, and CA produced using bio-based acetic acid (CA (bio-AA)), together with the contribution breakdown to climate change impact of material and energy inputs. d) Normalized impact indicators for each polymer, where 0 represents the lowest environmental impact and 1 the highest impact across each category. Absolute impact values for all categories and polymer candidates are provided in Table S10. Abbreviations: AA = acetic acid, non-renew = non-renewable, c = carcinogenic, nc = non-carcinogenic, ox = oxidation.

### 3.2.2. LCA comparison of fossil- and bio-based membranes

CA (bio-AA) was selected for membrane-level comparison, as it represented the most promising sustainable cellulose-based polymer alternative. For simplicity, cellulose-based membranes manufactured using CA (bio-AA) are hereafter referred to as regenerated cellulose (RC) membranes.

Figure 3 and Figure S5 show that, in more than 85 % of the simulations, RC membranes exhibit higher impact scores than PSF membranes across all categories. These results indicate that the higher environmental burden of RC membranes is not solely associated with replacing PSF with CA, which was already expected from the polymer-level comparison for categories such as climate change and eutrophication (Figure 2). Instead, additional burdens arise from the regeneration step required for cellulose-based membrane production, particularly due to ethanol use and its effect on wastewater streams. For example, moderate increases in median impact scores (13-17%) were observed for categories such as climate change, freshwater ecotoxicity, and human toxicity (carcinogenic). These categories are strongly driven by solvent impacts in PSF and RC membrane manufacturing (Figure 1b and Figure S4), resulting in comparatively smaller increases upon polymer substitution and membrane regeneration.

Larger increases in median impacts (24-36%) were observed for marine eutrophication, human toxicity (non-carcinogenic), and ozone depletion, which are primarily dominated by wastewater-related burdens (Figure 1b and Figure S4). In RC membranes, the additional presence of ethanol in wastewater streams increases the organic loading, leading to higher dilution requirements and wastewater burdens. Moreover, non-renewable energy demand showed the lowest relative increase, with median scores 8% higher for RC membranes compared to PSF. This smaller difference reflects the partial benefits associated with replacing fossil-derived PSF with CA (bio-AA), which partially offset the additional burdens mainly arising from the wastewater burden. On the other hand, RC membranes exhibited the largest increase in median land use score (84%). This increase is primarily associated with ethanol production, as its biomass-derived origin contributes approximately 20% of the total land use impact of RC membranes (Figure S4).

Replacing fossil-based polymers with bio-based alternatives does not necessarily translate into lower environmental impacts at the membrane level. Although CA (bio-AA) reduces dependence on fossil-derived feedstocks, the environmental benefits achieved at the material level in selected impact categories are partially offset by the additional regeneration step required to produce RC membranes. These results highlight that improvements observed at the material level do not necessarily propagate to the manufacturing system. Consequently, the sustainability of bio-based membranes must be evaluated across the entire manufacturing process rather than based solely on the origin or environmental profile of the polymer, as this may lead to misleading conclusions about the sustainability of the overall system.

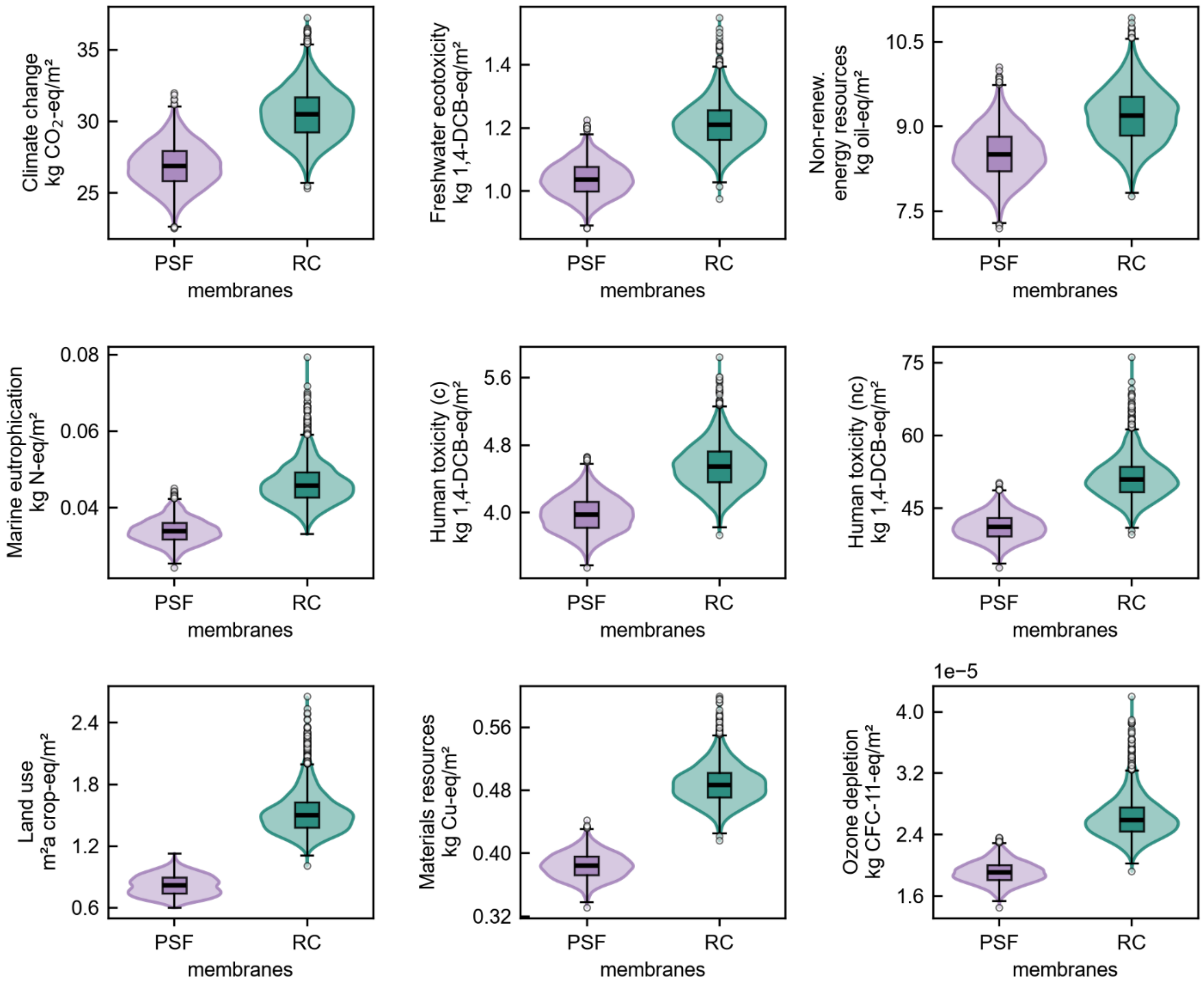

**Figure 3.** Violin plots overlaid with box plots comparing impact score distributions under uncertainty for PSF and regenerated cellulose (RC) membranes, reported per m² of membrane produced. Uncertainty was propagated using Latin Hypercube sampling (n = 3000) based on parameter distributions (Table S15-S16). Abbreviations: non-renew = non-renewable; c = carcinogenic; nc = non-carcinogenic.

## 3.3 Strategies to reduce environmental footprint in membrane manufacturing

### 3.3.1. Global sensitivity analysis

Global sensitivity analysis was used to identify the key parameters driving the environmental sustainability of membrane manufacturing and to determine opportunities for impact mitigation.

Figure 4 presents the Spearman correlation coefficients of the most influential parameters across PSF and RC membrane systems for five representative impact categories. The sensitivity analysis of all the other impact categories is provided in Figures S6-S7. Among all evaluated variables, the volume ratio: bs to ps parameter emerged as the dominant sustainability driver. This parameter quantifies the volume of bore solution (bs) consumed per volume of polymer solution (ps) prepared. Because the bore solution also contains solvent, increases in this ratio directly increase total solvent demand within the process. Consistent with this observation, the solvent fraction in the bore solution (solvent fraction, bs) and additive fraction were also identified as important influential parameters. Particularly, additive fraction is correlated with solvent demand, as both the additive and polymer fractions determine the remaining solvent content in the formulation. As a result, increasing the additive fraction reduces the solvent requirement in the polymer solution. Therefore, the resulting negative rank correlation observed in Figure 4 indicates that the environmental impact of the additive is primarily driven by its effect on solvent demand, rather than by its own upstream production burden.

Polymer demand, estimated through the polymer fraction, also shows a strong influence across the indicators. In the model, the polymer fraction (and its uncertainty range) is the same for both PSF and RC membranes; therefore, the higher influence observed in the RC system is driven by the higher environmental impacts associated with the CA polymer itself (Section 3.2.1), particularly in categories such as climate change, human toxicity, and land use. Additionally, in RC membranes, the regeneration step plays an important role in driving sustainability, with its influence governed by the NaOH concentration in the ethanol-based regeneration bath. Similar to the additive fraction, the negative Spearman correlation reflects the fact that, for a given NaOH requirement, higher NaOH concentrations reduce the volume of regeneration solution and consequently the ethanol demand, thereby lowering the associated environmental burdens. Overall, these results highlight that the environmental benefits of polymer substitution are only achievable when lower-impact polymers can be implemented without increasing polymer intensity or introducing additional material- and energy-intensive post-processing steps.

The rinsing stage emerges as a sustainability driver in membrane manufacturing, with water demand and temperature strongly impacting categories such as climate change, ecotoxicity, non-renewable energy use, and terrestrial acidification (Figure 4 and Figures S6 and S7). Both parameters directly affect the energy demand during rinsing, adding substantial environmental burdens across multiple impact categories. Finally, environmental categories such as human toxicity (non-carcinogenic) show strong sensitivity to the target COD, used to determine dilution requirements for NMP-containing effluents. This result further highlights the critical role of NMP as a key driver of environmental impacts, as its extensive use contributes to high COD loads in process effluents and consequently drives high dilution water demand and wastewater-related burdens.

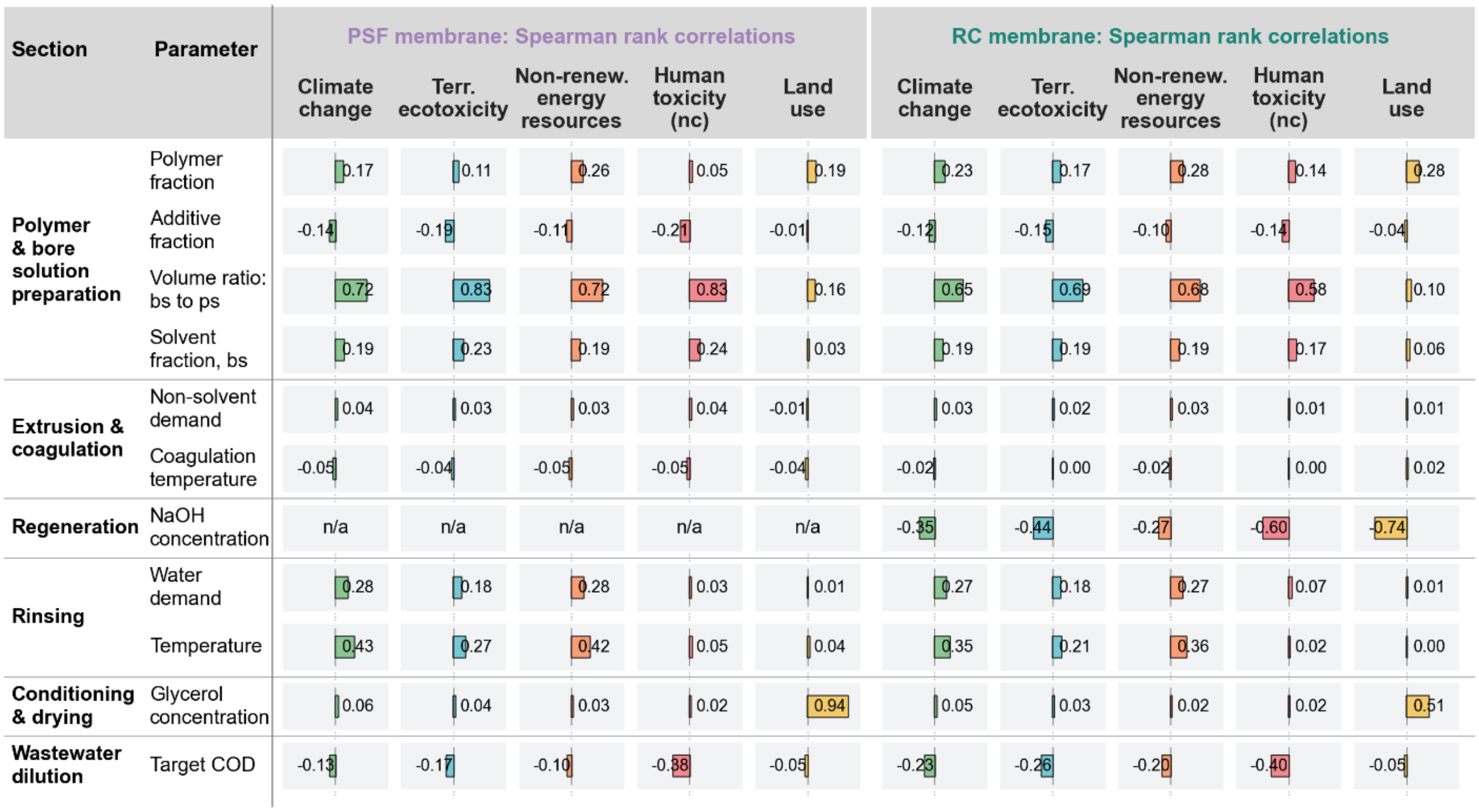

**Figure 4.** Sensitivity of environmental indicators to technological parameters in PSF and RC membrane manufacturing. Sensitivity was calculated using Spearman's rank correlations based on Latin Hypercube sampling (n = 3000 samples). Positive correlations indicate that increases in a parameter lead to higher environmental impacts, whereas negative correlations indicate that increases in a parameter reduce the impact score (i.e., improve environmental performance). Results are shown for five representative impact categories, while results for the remaining categories are provided in Figure S6. Abbreviations: bs = bore solution; ps = polymer solution; COD = chemical oxygen demand; terr = terrestrial; non-renew = non-renewable; nc = non-carcinogenic.

### 3.3.2. Solvent-related strategies

Due to the high solvent demand and the associated upstream and wastewater-related burdens, strategies focused on solvent replacement and solvent demand reductions could potentially translate into environmental benefits. For example, numerous alternatives to NMP have been proposed and often show environmental benefits at the membrane level when evaluated under unchanged membrane fabrication conditions. However, solvent substitution can alter polymer solution composition, bore liquid formulation, membrane yield, and downstream processing requirements[10,23]. Our sensitivity analysis highlights the importance of these effects, identifying solvent demand and various operating conditions as key sustainability drivers (Figure 4). Therefore, evaluating solvent substitution solely based on the environmental performance of the solvent may overlook important tradeoffs.

Another potential strategy for reducing solvent-related environmental burdens is solvent recovery and recycling from wastewater streams. Although solvent recovery has not been evaluated in previous membrane LCA studies, industrial practices suggest that solvent recycling may be implemented under certain conditions. The implementation of solvent recovery depends on several system-level factors, including plant scale, process integration (membrane-dedicated plant or within additional manufacturing lines), and solvent concentration in wastewater streams. Recovery efficiency and treatment requirements are strongly concentration-dependent[5]; therefore, the large water demand typically associated with membrane fabrication may become a limiting factor for solvent recovery by substantially diluting NMP-containing streams. Considering these factors, we evaluated a scenario in which NMP-containing wastewater streams are separately collected and treated through solvent recovery by distillation. Figure 5 compares the environmental impacts of PSF membrane manufacturing with and without solvent recycling under different water demand scenarios. Across the impact categories assessed, scenarios including solvent recycling exhibit greater sensitivity to water demand, as reflected in the sharp increase in impact scores relative to the baseline scenario without recycling.

For climate change and non-renewable energy use (Figure 5a and 5b), solvent recycling becomes environmentally favorable only when water demand during membrane fabrication is below 36 L $m^{-2}$ membrane (median score). At higher water demand values, the low NMP concentration (< 3 wt.%) in wastewater streams makes solvent recovery highly energy intensive, offsetting the environmental benefits associated with reduced solvent requirements. Similar trends were observed for terrestrial acidification, terrestrial ecotoxicity, and particulate matter (Figure S8). While energy demand is not the dominant contributor to these impact categories in the baseline system (Figure 1b), the increased steam consumption associated with solvent distillation leads to a substantial increase in their environmental impacts.

In contrast, categories such as human toxicity (non-carcinogenic) and marine eutrophication (Figure 5c and 5d), as well as marine ecotoxicity, freshwater ecotoxicity, freshwater eutrophication, ionizing radiation, ozone depletion, and water use (Figure S8), show environmental benefits from solvent recycling at water demand ranges between 100-300 L $m^{-2}$, which overlap with water demand values reported for industrial membrane manufacturing[5]. As discussed in Section 3.1 (Figure 1b), these categories are primarily dominated by solvent- and wastewater-related burdens. Consequently, reducing solvent demand simultaneously decreases upstream solvent impacts and the burdens associated with managing solvent-containing wastewater streams, allowing environmental benefits to emerge even under representative industrial water demand conditions.

Overall, reducing solvent-related impacts in membrane manufacturing will likely require a combination of lower-impact solvents and, where technically and economically feasible, solvent recovery strategies. However, the environmental benefits of both approaches strongly depend on maintaining favorable membrane fabrication conditions and minimizing process water demand, as additional solvent, energy, or wastewater burdens may offset potential sustainability gains.

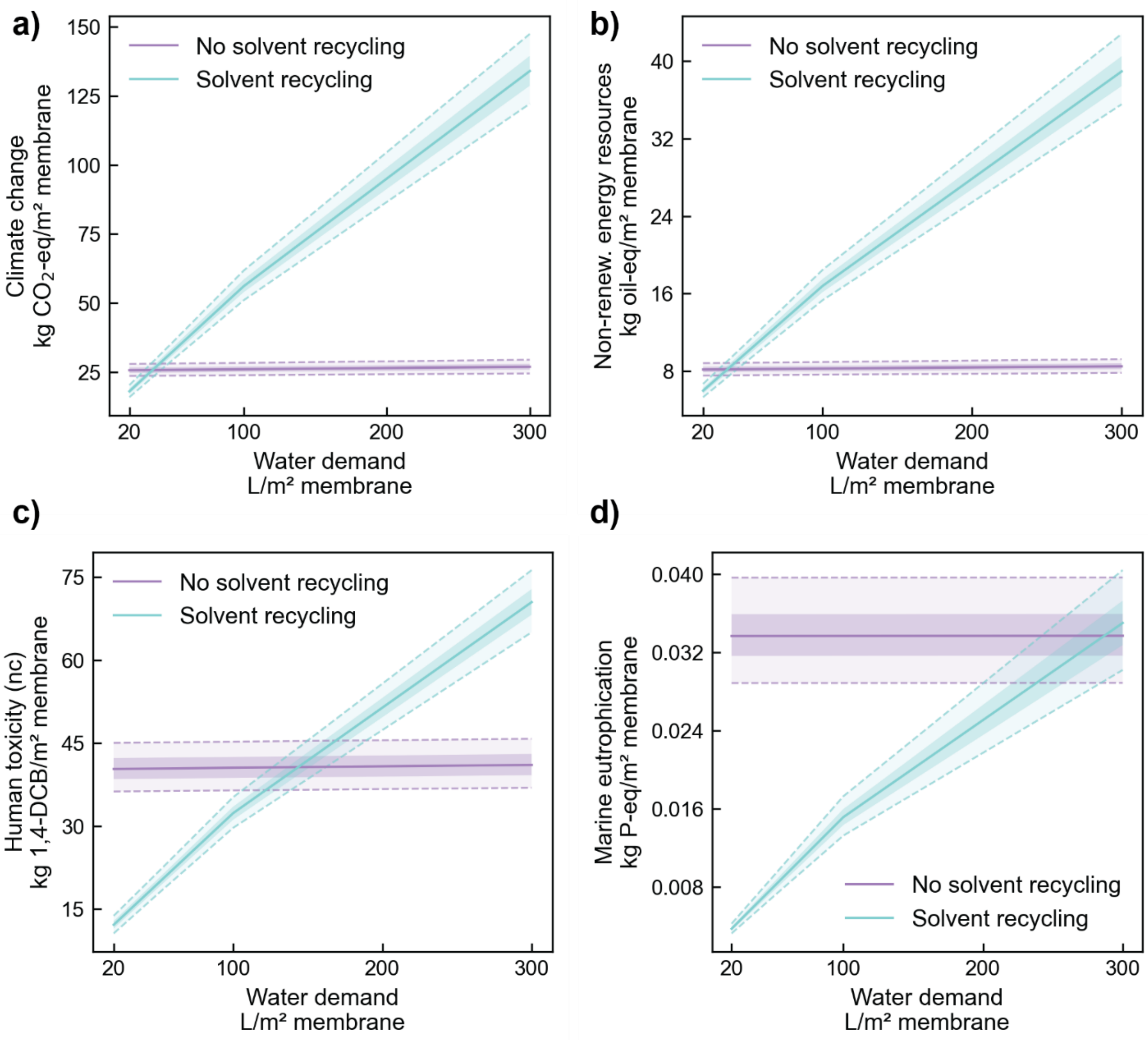

**Figure 5.** Comparison of environmental impacts for PSF membrane manufacturing with solvent recycling (sky blue) and without solvent recycling (purple) under varying water demand scenarios. Results are shown for: a) climate change, b) non-renew. energy resources, c) human toxicity (nc) and d) marine eutrophication. Uncertainty is represented by solid lines (medians), shaded regions (25th–75th percentiles), and dashed lines (5th–95th percentiles). Regions below the intersection point indicate environmentally favorable conditions for solvent recycling relative to the baseline scenario without recycling. Abbreviations: non-renew = non-renewable; nc = non-carcinogenic.

### 3.3.3. Rinsing and water-related strategies

Operating conditions during rinsing were identified as highly sensitive parameters driving sustainability, particularly temperature and water demand (Figure 4). The primary objective of rinsing is to remove residual solvent after the coagulation phase, and it is commonly performed using water baths at temperatures ranging from near-ambient to elevated conditions. While higher temperatures are generally expected to accelerate solvent removal, membrane studies lack information on the relationship between rinsing temperature and water demand. One reason may be that water demand is not typically treated as an optimized operating parameter, as it is often dictated by equipment design and manufacturing configuration. Nevertheless, our results identify water demand as a critical sustainability lever.

Figure 6 illustrates how optimizing rinsing temperature and total water demand could lead to improved sustainability. The lowest climate change and terrestrial ecotoxicity scores, approximately 25 kg $CO_2$-eq $m^{-2}$ and 200 kg 1,4-DCB-eq $m^{-2}$, respectively, are achieved at near-ambient temperatures across the entire range of water demand evaluated, as heating and cooling requirements are minimized. In contrast, both lower and higher operating temperatures increase impacts due to the additional energy required. Consequently, elevated temperatures become environmentally favorable only when they enable substantial reductions in water demand. For example, temperatures up to 50 °C can minimize climate change and terrestrial ecotoxicity if water consumption is reduced below 100 L $m^{-2}$. However, the combination of high temperatures (>50 °C) and large water demand (>200 L $m^{-2}$) represents the least favorable operating region, as the energy required to heat large water volumes adds substantial environmental burdens to the process.

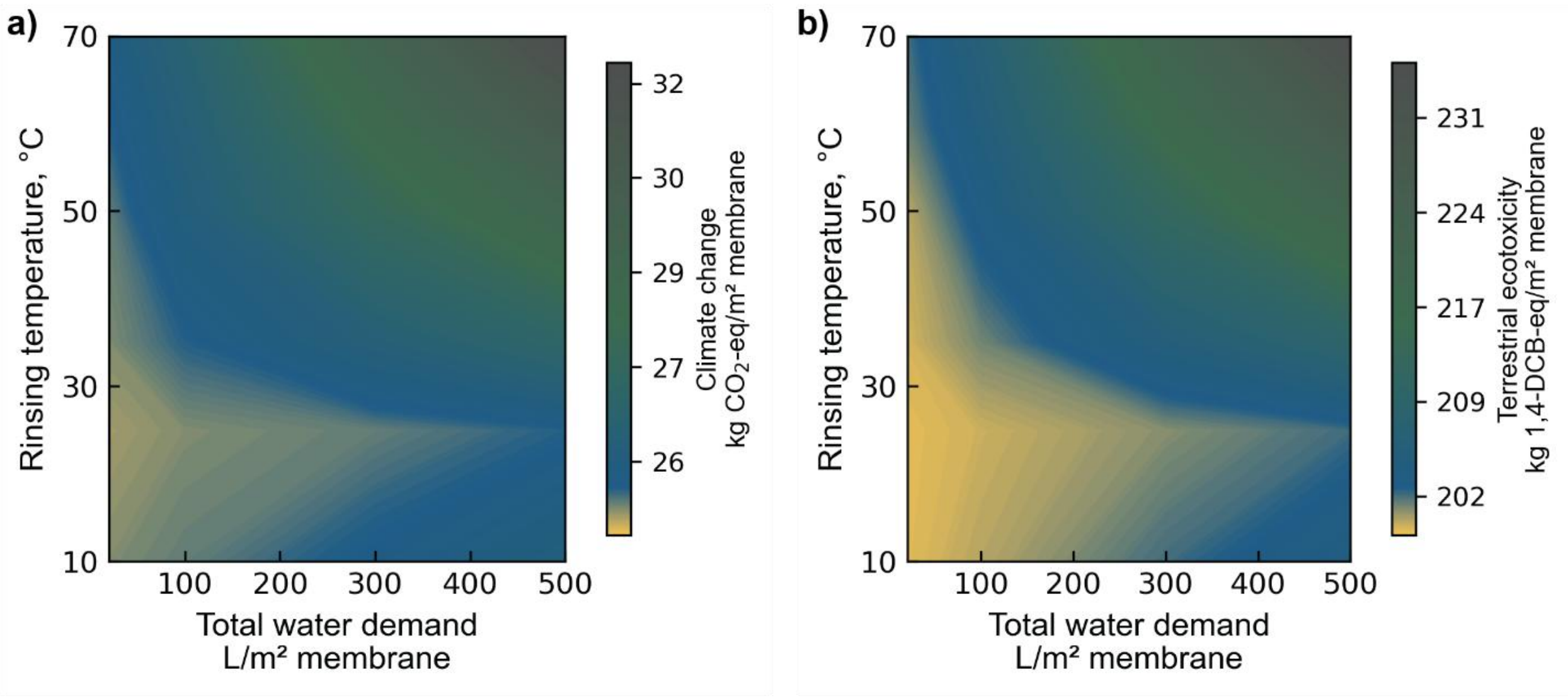

**Figure 6.** Environmental impacts for PSF membrane manufacturing as a function of rinsing temperature and total water demand. Results are shown for: a) climate change, and b) terrestrial ecotoxicity. Colors represent impact indicator values, with yellow and gray indicating the lowest and highest impacts, respectively. Results are presented for baseline membrane manufacturing conditions without solvent recycling.

These results suggest that increasing rinsing temperature as a strategy to accelerate solvent removal can provide environmental benefits only when accompanied by meaningful reductions in water demand. More broadly, they highlight the need to consider water consumption as a design variable in membrane manufacturing, as improvements in water-use efficiency may provide substantial sustainability gains while also enhancing the feasibility of solvent recovery strategies as discussed previously.

More broadly, additional opportunities for reducing water demand arise from alternative rinsing configurations (baths and washing technologies). A different strategy is alcohol-assisted rinsing, in which alcohol is added to the rinsing bath to modify solvent diffusion behavior[55]. Depending on the solvent-polymer and solvent-nonsolvent system, the addition of alcohol can increase the affinity of the residual solvent for the rinsing medium, accelerating solvent removal from the fibers[21,56,57], and potentially reducing both rinsing time and water demand. However, such a strategy also introduces additional material- and wastewater-related environmental burdens, creating potential trade-offs that should be carefully evaluated before being implemented. Alternatively, improvements can be achieved by optimizing the rinsing operation itself. Spray-based or other water-efficient rinsing systems may reduce water consumption while maintaining solvent removal efficiency, thereby lowering wastewater generation and associated environmental burdens.

### 3.4. Conclusions and future work

Achieving sustainable membrane manufacturing requires moving beyond material substitution alone and considering the full manufacturing system, including process design, resource consumption, and waste management. By integrating process simulation and life cycle assessment, this study identified polymer and bore solution preparation and wastewater dilution stages as the main environmental hotspots, demonstrating that both material selection and process efficiency govern the environmental performance of membrane manufacturing.

From a polymer substitution perspective, the use of bio-based polymers does not inherently translate into environmental benefits at either the material or membrane level, indicating that feedstock origin alone is insufficient to ensure environmental improvements. Overall, regenerated cellulose membranes exhibited higher environmental impacts than polysulfone membranes across all assessed impact categories. These results reflect both the chemical-intensive nature of cellulose modification pathways and the additional regeneration step needed to transform cellulose-derived polymers into membrane products. Consequently, future development of biobased polymers for membrane applications should prioritize not only lower-impact synthesis and modification pathways, but also materials that can be integrated within existing manufacturing routes with none or minimal additional processing requirements, while maintaining comparable membrane performance. From a process perspective, solvent recycling, water-use reduction, and rising optimization emerged as key opportunities for reducing environmental impacts. The environmental viability of solvent recycling was governed primarily by water demand, becoming favorable for several impact categories only below approximately 36 L $m^{-2}$, highlighting water consumption as the key factor enabling sustainable solvent recovery. More broadly, optimizing rinsing conditions to minimize water demand provides a practical pathway to simultaneously reduce wastewater generation and enhance the environmental benefits of solvent management.

Future work should extend solvent substitution assessments beyond solvent substitution to also evaluate the process modifications required for its implementation, as this study demonstrated that changes in solvent demand, operating conditions, and downstream operations can offset the benefits of lower-impact alternatives. Additionally, opportunities for material substitution should be expanded beyond the membrane polymer to include the cartridge housing. Replacing both membrane and housing materials with lower-impact polymers could further reduce life-cycle environmental impacts during manufacturing and end-of-life stages. Future assessments should therefore adopt a whole-life-cycle perspective to evaluate how material selection influences environmental performance from manufacturing through end-of-life.

AUTHOR INFORMATION

**Present Addresses**

†MilliporeSigma, Bedford, MA, 01730.

‡Department of Chemical Engineering, The Pennsylvania State University, University Park, PA, 16801; Institute of Energy and the Environment, The Pennsylvania State University, University Park, PA, 16801.

**Author Contributions**

The manuscript was written through contributions of all authors. All authors have given approval to the final version of the manuscript.

SUPPORTING INFORMATION

Supporting information: Additional text, data, tables, and figures supporting the results and discussion presented in this article (PDF).

The code used in this study is openly available at GitHub at https://github.com/ElizabethAigaje/Polymeric_membrane_manufacturing

ACKNOWLEDGMENT

Support for this work was provided through the Membrane Applications Science and Technology (MAST) Center, funded by grant number 2310832 from the National Science Foundation's IUCRC program.